\documentstyle[prl,aps]{revtex}
\twocolumn
\newcommand{\BEQ}{\begin{equation}}                                           
\newcommand{\EEQ}{\end{equation}}
\newcommand{\BEA}{\begin{eqnarray}}                              
\newcommand{\EEA}{\end{eqnarray}}

\renewcommand{\d}{{\rm d}}
\newcommand{\I}{{\cal I}}
 \begin{document}
 \draft
\title{Thermodynamic description of a dynamical glassy transition}
\author{Th.~M.~Nieuwenhuizen}
\address{ Van der Waals-Zeeman Instituut, Universiteit van Amsterdam\\ 
	 Valckenierstraat 65, 1018 XE Amsterdam, The Netherlands\\}
 \date{21-8-1997} 
\maketitle
\begin{abstract}
For the dynamical glassy transition in the $p$-spin mean field 
spin glass model a thermodynamic description is given.
The often considered marginal  states are not the 
relevant ones for this purpose.
This leads to consider a cooling experiment on exponential
timescales, where lower states are accessed. 
The very slow configurational modes are at quasi-equilibrium at
an effective temperature. 
A system independent law is derived that expresses their
contribution to the specific heat. $t/t_w$-scaling in the aging
regime of two-time quantities is explained.
\end{abstract}
\narrowtext

The structural glass transition occurs only in an idealized adiabatic cooling
procedure at the Kauzmann temperature $T_K$. In a realistic experiment with 
finite cooling rate a gradual freezing transition takes place in a small
 interval centered around a higher temperature $T_f$, that depends on the 
cooling rate.  Although the freezing is not a sharp thermodynamic 
transition, there can be some 15 decades in time involved, 
from picoseconds to many hours. By extrapolation from the high and
low temperature sides, one may define
jumps in quantities such as the specific heat and the compressibility. 
It has been pointed out by J\"ackle ~\cite{Jackle} and Palmer ~\cite{Palmer}
that the freezing transition can be described as
a smeared thermodynamic transition, on a thermodynamic basis, 
with ensemble averages replacing time averages. The observation time sets
the time scale that separates ``fast'' processes (time scale less that 
the observation time) from ``slow'' ones. The latter are essentially frozen.
 
Upon cooling, a liquid freezes dynamically in a glassy
state with extensively smaller entropy.
The free energy of the glassy state then is much larger, and
 it is not obvious why the system can get captured in such a state.
 The point is that the condensed system has lost  the 
 entropy of selecting one out of the many 
 equivalent states; this part of the entropy is called the 
 {\it complexity}, {\it configurational entropy}, or {\it information entropy}
  ${\cal I}$.~\cite{Jackle}~\cite{Palmer}
This can be understood as follows.
For long times the system is stuck in states with long (but finite) lifetime.
These states are called  ``components'' by Palmer~\cite{Palmer} and
`states' or `TAP-states' in spin glass theory.
When the Gibbs free energy $F_{\bar a}$ of the relevant state $\bar a$ has a 
large degeneracy ${\cal N}_{\bar a}\equiv \exp({\cal I}_{\bar a})$, the
 partition sum yields $Z=\sum_a\exp(-\beta F_a)\approx 
{\cal N}_{\bar a}\exp(-\beta F_{\bar a})$, so 
 $F=F_{\bar a}-T{\cal I}_{\bar a}$ 
is the total free energy of the system. 
 The entropy loss arises when the system chooses the state to condense into,
 since from then on only that single state is observed.~\cite{quantummeas}
 As the total entropy $S=S_{\bar a}+{\cal I}_{\bar a}$ is continuous, 
 so is the total free energy. 
 
The aim of the present work is to analyze this thermodynamical picture
of the dynamical freezing transition in a well understood mean field model.
The difficulty is to extract the information that is not due to the
mean field limit.
The gain is that we find strong constraints satisfied by the dynamics,
without having solved the it.
We first discuss the present status of marginal replica theory, 
its relation
with dynamics and its fundamental flaws. Then we shall propose
a solution to these paradoxes.

 We consider the mean  field $p$-spin interaction spin glass model
of  $N$ coupled spins in a field with Hamiltonian 
\begin{equation}\label{Ham=}
{\cal H}=-\sum_{i_1<i_2<\cdots<i_p} J_{i_1 i_2\cdots i_p}
S_{i_1}S_{i_2}\cdots S_{i_p}-H\sum_iS_i
\end{equation}
The independent Gaussian random couplings $J_{i_1 i_2\cdots i_p}$
have average zero
and variance $J^2p!/2N^{p-1}$. The spins are subject to the spherical
condition $\sum_i S_i^2=N$.

This model has  a close analogy with
models for the structural glass transition. ~\cite{KTW}
On a static level  there occurs a transition to a state with 
one step replica symmetry breaking (1RSB) at a temperature  
$T_K$, comparable to the ideal glass transition in an  
 adiabatic cooling experiment.
The 1RSB replica calculation involves parameters three parameters.
The overlap of spin configurations in two states can be equal 
the Edwards-Anderson parameter or self-overlap $q_1$, or have the 
smaller value  $q_0$ $(=0$ for $H=0)$; these values occur
with probability $1-x$, and $x$, respectively.  The free energy reads
\begin{eqnarray}\label{bFCS}
\frac{ F}{N}&=&-\frac{\beta J^2}{4}(1-\xi q_1^p-xq_0^p)
-\frac{\beta H^2}{2}Q\\ 
&-&\frac{T}{2x}\log Q 
+\frac{T\xi}{2x}\log(1-q_1)
-\frac{Tq_0}{2Q} 
\nonumber\end{eqnarray}
where $\xi=1-x$ and $Q=1-\xi q_1-xq_0$ . Unless stated differently,
 we shall take $H=0$. 
$q_0$ and $q_1$ are determined by optimizing $F$. 
For $x$ the situation is not unique, but depends on the time-scale
considered. Setting $\partial F/\partial x=0$ 
yields the static phase transition at $T\equiv T_K$.~\cite{Istat} 
When considering the Langevin dynamics of this model, one may
derive dynamical equations for correlation and response functions 
by taking first first $N\to\infty$.~\cite{CHS} Solving for large $t$ 
leads to a sharp phase transition at larger temperature 
 $T_{A}=J \{p(p-2)^{p-2}/2(p-1)^{p-1}\}^{1/2}$, and a different form
$x(T)$. This dynamical value for $x$ can be simply 
rederived from a replica calculation 
in which the ``replicon'' or ``ergodon'' ~\cite{Nqsg}
fluctuation mode is taken to be massless.   
This leads to the marginality criterion 
$p(p-1)\beta^2 J^2q_1^{p-2}(1-q_1)^2/2=1$.
At the dynamical transition there is  a sharp
jump in the specific heat.~\cite{CHS}

To discuss the situation,
we must first consider the TAP states. 
A state is labeled by $a$ and has local magnetizations
$m_i^a=\langle S_i\rangle^a$. Its free energy 
$F_a(T)$ is a thermodynamic potential that determines the internal energy
and the entropy by its derivatives. In the present model  
$F_a=F_{TAP}(m_i^a)$ is known explicitly. It is a minimum of the
``TAP'' free energy functional~\cite{Rieger}~\cite{KPV}~\cite{CStap}
\begin{eqnarray} \label{FTAP=}
&F&_{TAP}(m_i)=
-\sum_{i_1<\cdots<i_p}J_{i_1 
\cdots i_p}m_{i_1}
\cdots m_{i_p} -H\sum_i m_i \nonumber\\
&-&\frac{NT}{2}\log(1-q)
-\frac{N\beta J^2}{4}(1+(p-1)q^p-pq^{p-1})
\end{eqnarray}
where $q=(1/N)\sum_i m_i^2$ is the self-overlap. Below we shall argue that
the commonly used Gibbs weight, $p_a=\exp(-\beta F_a(T))/Z$ 
is the relevant one. Given the $p_a$'s one can define the 
component averages such as $\overline F=\sum_a p_aF_a$, 
$\overline U=\sum_ap_aU_a$, $\overline C=\sum_a p_aC_a$, and even the
complexity ~\cite{Jackle}~\cite{Palmer}
${\cal I}=-\sum_a p_a \ln{p_a}$. For observables the direct evaluation from
the ordinary partition sum should coincide with the outcome of the 
TAP-analysis: $U=\overline{U}$, $M=\overline{M}$. 
They need not be derivatives of $\overline F$.

For all $T<T_A$ the mode-coupling equations 
a massless ergodon, responsible for aging phenomena.
We recently assumed that this is a very general phenomenon. 
will automatically  get trapped in a state with 
diverging time scale, whenever present.
The marginal replica free energy has the form
\BEQ\label{FIcx}
F={\overline F}-\frac{T\I_c}{x}
\EEQ
where 
\begin{equation} \label{I=}
{\cal I}_c=N\left(\frac{1}{2}\log(p-1)+\frac{2}{p}-1\right)
\qquad{\rm (marginality)}
\end{equation}
is the complexity of the marginal states.
Below $T_A$ the free energy lies below the one of the paramagnet and
has a larger slope. 
This would naively imply a latent heat.

There is, however, another prediction for the free energy.~\cite{CHS}
It involves the internal energy and an  entropy obtained by 
integrating $(1/T)\partial U/\partial T$ from a temperature in the glassy
phase up to some large temperature.
The resulting ``experimental'' glassy free energy~\cite{CHS} 
\BEQ\label{FIc}
F_{exp}={\overline F}-T\I_c
\EEQ
exceeds the paramagnetic free energy 
quadratically and is by construction a thermodynamic potential. 
It was reproduced by 
analysis of the TAP states.~\cite{CStap} 

The difference between (\ref{FIcx}) and (\ref{FIc})
 led us to question fundamentally 
the validity of replica calculations for dynamical 1RSB transitions.
Our aim was to find the meaning of the logarithm of the 
dynamical replica free energy (eq. (\ref{bFCS}) 
with $\partial F/\partial q_0=\partial F/\partial q_1=0$
but with $\partial F/\partial x\neq 0$).

By doing the full thermodynamic analysis of the 
TAP-partition sum $\sum_a\exp(-\beta F_a)$ at $H=0$, 
we found that that the replica free energy (\ref{FIcx})
is reproduced.~\cite{Ncompl} 
As the glassy free energy lies below the continuation of the
paramagnetic one, we considered this as proof that the complexity
is the driving force for the dynamical phase transition.~\cite{Ncompl}
In the doing the analysis we realized that the calculation of 
the dynamical complexity cannot be separated from the
calculation of the free energy, and that replica symmetry breaking 
is essential. The problem boils down to
a replicated TAP free energy that has 6 replicated order parameters.
We have now extended this analysis to $H>0$. 
For 1RSB with a common breakpoint $\tilde x$, each replica order parameter
now brings 3 parameters. We thus obtain an
optimization problem in 18 variables, that was solved partly using Maple.
We have verified that the total free energy, the internal energy and the 
magnetization, calculated within the TAP-approach, coincide with their 
replica values. For the magnetization this is particularly satisfying, 
as $M_a$ of a given marginal TAP-state is temperature independent. 
(It then holds that $\partial M_a/\partial T =\partial S_a/\partial H=0$. 
Nevertheless, the component average $\overline {M}=\sum_ap_aM_a$ is 
 temperature and field dependent, and equal to the replica value
$M=\beta H(1-\xi q_1-x q_0)$).

For interpreting eq. (\ref{FIcx}) one might be tempted to
consider $\I_c/x$ as the full complexity.  Since $x\to 0$ for $T\to 0$,
this is hard to explain on a physical basis, however.
The $1/x$ dependence in (\ref{FIcx}) does not disappear after 
the quantization
of the spherical model, recently proposed by us.~\cite{Nqsm,Nqsg}
Analysis of the equations for quantized spherical spins, or for Ising spins, 
learns us that, though $T_A$ shifts, the
term $T\I_c/x(T)$ survives for marginal states.  

The other interpretation of eq. (\ref{FIcx})
is that $T_e\equiv T/x$ is an
effective temperature at which the slow processes leading
to  $\I_c$ are in quasi-equilibrium.
 This interpretation is promising, since
the same effective temperature shows up in the fluctuation-dissipation
relation in the aging regime of the mode-coupling equations.~\cite{CuKu}

There remain some paradoxes connected to the marginal states.
For large $p$ one has $\I_c\sim (N/2)\log p$, which
(for quantized spherical spins or for Ising spins)
already exceeds the total entropy available. 
This shows that {\it the dynamical transition at $T_A$
has no thermodynamic counterpart in short range systems}, 
at least for large enough $p$. 
In fact, $T_A$ may be identified with the
critical temperature of mode-coupling theory, 
which lies well above the
freezing temperature $T_f$.
Even more cumbersome is that for marginal states with 
$H\neq 0$ we have shown violation of the inequality 
$C\ge\overline{C}$~\cite{Ncompl}. This says that  
marginal states are intermediate-time states, from which the
system must escape at longer times.

The implication of these arguments is rather dramatic:
 for comparing with realistic short range systems 
{\it we must abandon the assumption that the marginal states
are the physically relevant states.}
Let us see how this could happen.  
Marginality arises automatically in the dynamical equations 
after taking first the limit $N\to\infty$ and then $t\to\infty$.
This order of limits, however, prevents all activated processes, 
which in the mean field model would need a time $\sim \exp(N)$. 
As a result, all dynamics is confined to the highest TAP-states,
which are marginal. The lower states are never reached.
In a  realistic short range glassy system, however, 
there is no sharp distinction between slow and
activated processes, and the latter can certainly not be omitted. 
So in a realistic experiment we do expect to reach lower states.

In order to compare to realistic systems and to avoid thermodynamic
paradoxes, we propose another look at the mean field system.
We consider the system at fixed large $N$ 
under such conditions that a range of 
TAP-states below the highest (marginal) ones are accessed.
We thus focus our attention on the free energy
of the state, and not so much on the timescale needed to reach it.
At $H=0$ the free energy of the TAP states can be  
characterized by a parameter $\eta$ ($\eta_{st}\le\eta\le 1)$, 
that enters the condition 
\BEQ
\beta ^2 p(p-1) q^{p-2}(1-q)^2/2=\eta.
\EEQ
For $\eta=1$ this would be the marginality condition, while
the static equations can also be put in this form ~\cite{CS}
with $\eta=\eta_{st}<1$ independent of $T$. 
One now finds the breakpoint $x=(1-q)(p-1-\eta)/q\eta$, so
choosing $\eta$ between $\eta_{st}$ and $1$ 
can alternatively be seen as a way of fixing the
mysterious parameter $x$, for which no obvious criterion was present.
In the present approach it is directly related to the time scale 
$t_\eta\equiv\exp(N\tau_\eta)$ at which the $\eta$-states are reached. 
Unfortunately, the precise relation between the logarithmic 
time variable $\tau_\eta$ and  the lowest reachable TAP free energy 
at that scale, $F_\eta$, is unknown.

We assume that at a given exponential timescale $\tau_\eta$
the dominant
states are determined by a saddle point. Such a behavior was
seen in a solvable glassy transition in a directed polymer model,
recently proposed by us.~\cite{Ndirpol} This allows to restrict
the discussion to TAP-states with a common free energy.

We have verified that
the equivalence between the replica and the TAP analysis also holds 
for $\eta<1$. The $\eta$-states have  complexity
\BEQ
\I_c(\eta)=\frac{N}{2}\{\log\frac{p-1}{\eta}
-\frac{(p-1-\eta)(\eta+1)}{p\eta}\}
\EEQ
As $\eta$ decreases from $1$ to $\eta_{st}$,   
$\I_c(\eta)$ goes down from the
marginal value $\I_c(1)$ to the static value  $\I_c(\eta_{st})=0$. 
A sensible cooling experiment must have a temperature
dependent value of $\eta$, such that
$\eta\to\eta_{st}$ fast enough for $T\to 0$.
In hindsight this is just what is needed to describe a 
realistic cooling experiment in the mean field $p$-spin model.  
Let us  assume to have some logarithmically slow 
cooling traject $T(t)$ where also $\eta=\eta(t)$ depends on time
in a still unknown, but determined  fashion. 
We can then eliminate $t$ and
construct the function $T(\eta)$ (and its inverse $\eta(T)$) 
that characterizes our cooling traject.
 A dynamical freezing transition will occur when the lowest state reached
at time $t$ freezes at the temperature $T(t)$.
This occurs at $T_f=T(\eta_f)$ 
where $\eta_f$ is the solution of 
\BEQ 
T(\eta)=J\left\{
\frac{p\eta(p-1-\eta)^{p-2}}{2(p-1)^{p-1}}\right \}^{1/2}
\EEQ
As $T(t)$ and $\eta(t)$ will depend on
the cooling procedure, we expect $T(\eta)$ to do the same, 
so that $T_f$ will not be universal but depend on the specific traject. 

Also the assumption of Gibbs weights is now justifyable. TAP states with
free energy larger than or equal to the ones fixed by $\eta$ are now 
effectively in thermodynamic equilibrium, and may be described by 
the Gibbs weight. Lower ones play no role anyhow.

The paradoxes related to marginal states, signaled here,
occur much more widely.
At present it is a whole field of research to consider dynamics of  
mean-field models by first taking the mean field limit and
then considering large times. 	
Though the approach has relevance for short and intermediate
time dynamics, its long time regime is a result of 
``squeezing the system into marginal states'', which has no bearing on
the long time relaxation of short range models.
This is already expressed by 
the unique sharp dynamical transition temperature $T_A$, found in 
the dynamical approach. It
 disappears on exponential time scales,
and is replaced by a cooling-rate dependent freezing temperature.

The basic object in our approach is the complexity.
An important question is whether it can be measured in the 
glassy phase. If so, it should be related to the specific heat
or the temporal energy fluctuations.
When monitoring the internal energy as function of time,
as is easily done in a numerical experiment, one obtains essentially a noisy
telegraph signal. Each plateau describes trapping in a TAP-state
for some definite time. The variance of the noise in the internal energy
on this plateau is equal to $T^2\overline C$=$T^2\sum p_aC_a$
=$\sum_a\langle(\delta U_a)^2\rangle$=$T^2\sum p_a \d U_a/\d T$.
 From time to time the system  moves to another TAP-state, causing 
additional noise. The variance of the total noise equals $T^2C$, 
and it should exceed $T^2\overline C$.~\cite{Palmer}
In ref. \cite{Ncompl} it was pointed out that there can be an extensive 
difference between the specific heat $C=\d U/\d T=
\d {\overline U}/\d T=\sum_a \d (p_aU_a)/\d T$ and
the component average energy fluctuations
$\overline C$$=$$T\d \overline S/\d T$
We can now  consider their difference at $H=0$ in a cooling experiment
of the type introduced above. We find the {\it excess specific heat}
\BEQ \label{DeltaC=}
\Delta C\equiv C-\overline{C}
=N\frac{q(1-\eta)}{2p\eta(1-q)}\,\,\frac{\d\eta}{\d T}
\EEQ 
Using this we can easily verify the fundamental relation
\BEQ\label{fund}
\frac{\d}{\d T}\I_c(T)
=\frac{1}{T} x(T)\Delta C(T)
\EEQ
It expresses the complexity in terms of 
measurable quantities, namely
the excess specific heat and the ergodicity-breaking parameter $x$. 
Equation (\ref{fund}) holds equally well for $H\neq 0$ but fixed. 
In the generalization of (\ref{DeltaC=}) one now encounters parameters 
$\eta(T;H)$ and its derivative 
$\partial\eta/\partial T$. The proof of eq. (\ref{fund})
for $H\neq 0$ is lengthy but could be 
 verified using Maple.~\cite{negDC}

The complexity can also be measured along the transition 
line in the $(T,H)$-plane. This issue is related with the Ehrenfest
relations, and has been discussed elsewhere.~\cite{NEhrenf}

With $T_e=T/x$ eq. (\ref{fund}) can also be written as
\BEQ\label{2ndlaw'}
\frac{\d U}{\d T}=T\frac{\d {\overline S}}{\d T}+T_e\frac{\d \I_c}{\d T}
\EEQ

Our interpretation of the replica results leads to an effective
temperature $T_e(t)=T(t)/x(t)$. 
The slowest active modes are at quasi-equilibrium
at this effective temperature. This explains why
$T_e$ also shows up
in the fluctuation-dissipation relation.~\cite{CuKu},\cite{CKP}
As they set the slowest time-scale, they must also dominate
the dynamical free energy. This is why the change of 
structural modes $\d\I_c/\d T$ in eq. (\ref{2ndlaw'}) 
has prefactor $T_e$.  

In numerics on the fluctuation-dissipation relation in
spin glasses and even binary soft spheres ~\cite{Parisi} it has 
been observed that the factor $x$ is linear
in $T$. Let us give a simple explanation for that behavior. 
As the effective temperature must exceed the Kauzmann temperature,
we can estimate $T_e(T)\approx $ const, implying
indeed $x=T/T_e\sim T$.

If we quench the system from high temperatures 
and let it evolve freely during a waiting time $t_w$, 
$T_e$ will be set by equating the equilibrium relaxation time $\tau_{eq}$ 
at the instantaneous  $T_e$ to $t_w$: $\tau_{eq}(T_e)=t_w$. 
Naively, one expects two-time quantities in dynamics
to be a function of $(t-t_w)/\tau_{eq}(T_e)=t/t_w-1$, explaining
immediately the often observed $t/t_w$ scaling in the aging 
regime. There could be logarithmic corrections
to this behavior.

Replacing $x\to T/T_e$ we can write eqs (\ref{bFCS})
and (\ref{FIcx}) as
\BEQ\label{FTTe} 
F(T,T_e)=U-T{\overline S}-T_e\I_c
\EEQ
It is a dynamical free energy
determining ${\overline S}=-\partial F/\partial T$ and 
$\I_c=-\partial F/\partial T_e$, while 
$U=F+T{\overline S}+T_e\I_c$. 
The system-independent laws (\ref{2ndlaw'}) and (\ref{FTTe}) 
are the cornerstone for our
thermodynamic description of the dynamical glassy transition, 
and expected to be valid in general.

We have verified eq. (\ref{2ndlaw'}) for a cooling
procedure in the Ising chain with Glauber dynamics. At $T=0$ it is
a coarsening problem of alternating up and down clusters of
average length $\xi(t)=\sqrt{4\pi t}$ and energy $U(t)=NJ(-1+2/\xi)$.
counted by standard manners. Its logarithm reads
$\I_c=N(1+\ln\xi)/\xi$. From the internal energy one may introduce 
an effective temperature: $1/\xi=e^{-2\beta_e J}$$\to$
$T_e=2J/\ln\xi(t)$. The same value would
follow when defining $T_e$  from the complexity $\I_c$.
Both results can thus be combined in terms of a  
dynamical free energy $F=-NT_e\ln(2\cosh\beta_eJ)$. This is a special
case of eq. (\ref{FTTe}) with ${\overline S}=0$.

As above,  we then
consider a cooling experiment where
$T(t)=2x(t)J/\ln\xi(t)$ with a smooth decreasing function $x(t)$. 
In the initial regime $x(t)>1$ the system
will achieve equilibrium at the instantaneous temperature $T(t)$.
For $x(t)<1$ this will not happen. The system falls out of
equilibrium and behaves as at $T=0$: it is at quasi-equilibrium at 
$T_e(t)=T(t)/x(t)$. The freezing transition
occurs around $T_f=T(t_f)$ where $x(t_f)=1$. In the frozen
phase eqs. (\ref{2ndlaw'},\ref{FTTe})
are valid with ${\overline S}=0$.
At $T=0$ (and for all  $T<T_f$) the on-site 
correlation function is found to be a scaling function of $t/t_w$,

Summarizing, we have shown that a thermodynamic description of a
dynamical freezing transition can be given. ~\cite{Ritort}
We have been led  to discard the whole issue of marginal states. 
For comparing with realistic short range systems,
we consider the mean field system at exponential time scales, where
lower states are accessed. They contribute to the partition sum
of Gibbs weights over dynamically relevant states. This approach
naturally leads to slow cooling procedures where a dynamical freezing
transition occurs at a tunable temperature.
This dynamical transition is described by a free energy that
depends on the real temperature and on an effective temperature.

\acknowledgments
The author thanks H.F.M. Knops, F. Ritort, and M. van Zuijlen 
for discussion. 

\end{document}